\documentclass[aps,preprint,tightenlines,showkeys,nofootinbib]{revtex4}

\usepackage{graphicx}  
\usepackage{bm}  
\usepackage{amsmath}
\newcommand{\beq}{\begin{equation}}
\newcommand{\eeq}{\end{equation}}
\newcommand{\bea}{\begin{eqnarray}}
\newcommand{\eea}{\end{eqnarray}} 
\newcommand{\beqa}{\begin{eqnarray}}
\newcommand{\eeqa}{\end{eqnarray}}

\newcommand{\sing}{^1{\rm S}_0}
\newcommand{\trip}{^3{\rm S}_1}

\newcommand{\mpic}{m_\pi^{crit}}
\newcommand{\eftnopi}{\mbox{EFT$(\not\!\pi)$\:}}
\newcommand{\eftnopinew}{\mbox{EFT$(\not\!\pi)$}}

\begin{document}
\title{Pion-mass dependence of three-nucleon observables}
\author{H.-W. Hammer}\email{hammer@itkp.uni-bonn.de}
\affiliation{Helmholtz-Institut f\"ur Strahlen- und Kernphysik (Theorie),
Universit\"at Bonn, D-53115 Bonn, Germany}
\author{D.R. Phillips}\email{phillips@phy.ohiou.edu}
\affiliation{Department of Physics and Astronomy,
Ohio University, Athens, OH 45701, USA\\}
\author{L. Platter}\email{lplatter@phy.ohiou.edu}
\affiliation{Department of Physics and Astronomy,
Ohio University, Athens, OH 45701, USA\\}

\date{\today}
\begin{abstract}
We use an effective field theory (EFT) which contains
only short-range interactions to study the dependence of a variety of
three-nucleon observables on the pion mass. The pion-mass dependence
of input quantities in our ``pionless'' EFT is obtained from a recent
chiral EFT calculation.  To the order we work at, these quantities are
the ${}^1$S$_0$ scattering length and effective range, the deuteron
binding energy, the ${}^3$S$_1$ effective range, and the binding
energy of one three-nucleon bound state. The chiral EFT input we use
has the inverse ${}^3$S$_1$ and ${}^1$S$_0$ scattering lengths
vanishing at $\mpic=197.8577$ MeV. At this ``critical'' pion mass, the
triton has infinitely many excited states with an accumulation point
at the three-nucleon threshold. We compute the binding energies of
these states up to next-to-next-to-leading order in the pionless EFT
and study the convergence pattern of the EFT in the vicinity of the
critical pion mass.  Furthermore, we use the pionless EFT to predict
how doublet and quartet $nd$ scattering lengths depend on $m_\pi$ in
the region between the physical pion mass and $m_\pi=\mpic$ .
\end{abstract}


\maketitle

\section{Introduction}

The NPLQCD collaboration recently computed $NN$ correlation functions
in QCD by using Monte Carlo techniques to evaluate the QCD path
integral on a discrete Euclidean space-time lattice
\cite{Beane:2006mx}.  This provided the first calculation of nuclear
physics quantities from full QCD: the $NN$ scattering lengths in the
${}^3$S$_1$ and ${}^1$S$_0$ channel.  However, the NPLQCD computation
was performed at quark masses which are significantly larger than
those in the physical world. This leaves us with the challenge of
understanding how observables such as the $NN$
scattering lengths depend on parameters of QCD such as $m_u$ and
$m_d$.

This challenge can be addressed using effective field theories
(EFTs). EFTs allow the calculation of physical observables as an
expansion in a small parameter that is a ratio of physical scales.  In
this paper, we will discuss two different EFTs. The first describes the
low-energy sector of QCD by exploiting QCD's approximate chiral
symmetry \cite{Bernard:2006gx}.
This chiral EFT ($\chi$EFT) is a powerful tool for analyzing the
properties of hadronic systems at low energies in a systematic and
model-independent way. It is formulated in an expansion around the
chiral limit of QCD which governs low-energy hadron structure and
dynamics. In $\chi$EFTs, the quark-mass dependence of operators in
the effective Lagrangian is included explicitly. Loops also generate
non-analytic dependence on the quark mass, but their effects can be
computed in a controlled way, since they can be obtained up to a given
order in the expansion parameter ${m_q}/{\Lambda_\chi}$, where
$\Lambda_\chi \sim m_\rho$ is the scale of chiral symmetry breaking in
QCD.  Over the past 15 years, considerable progress has been made in
understanding the structure of the nuclear force in this framework
\cite{Weinberg:1991um,Beane:2000fx,Bedaque:2002mn,Epelbaum:2005pn} but
some questions regarding the power counting
remain~\cite{Nogga:2005hy,Valderrama:2005wv,Birse:2005um,Epelbaum:2006pt,Mondejar:2006yu}.

The quark-mass dependence of the chiral nucleon-nucleon ($NN$)
interaction was studied in
Refs.~\cite{Beane:2001bc,Beane:2002xf,Epelbaum:2002gb}.  These studies
found that the inverse scattering lengths in the $\trip$--$^3{\rm
  D}_1$ and $\sing$ channels may both vanish if one extrapolates away from
the physical limit to slightly larger quark masses.\footnote{Due to
  the nuclear tensor force, the $\trip$ and $^3{\rm D}_1$ channels are
  coupled. This mixing is included in the $\chi$EFT calculations
  while it appears as a higher-order effect in the pionless EFT
  discussed below. For simplicity, we will only refer to the $\trip$
  and $\sing$ partial waves in the following.}  Subsequently, it was
pointed out that QCD is close to the critical trajectory for an
infrared RG limit cycle in the 3-nucleon sector. This led to the
conjecture that QCD could be tuned to lie precisely on the critical
trajectory through small changes in the up and down quark masses away from
their physical values \cite{Braaten:2003eu}.

In the vicinity of this critical trajectory another EFT becomes
useful. In this EFT observables are calculated as an expansion in
powers of $R/|a|$, where $R$ is the range of the two-body potential,
and $a$ is the two-particle scattering length. In nuclear physics this
is the ``pionless'' EFT;
\eftnopi~\cite{Weinberg:1991um,Kaplan:1998we,vanKolck:1997,Bedaque:1997qi,Gegelia:1998,Birse:1998}.
\eftnopi has been used to compute a number of $NN$ system observables
as a function of $R/|a|$~\cite{Beane:2000fx,Bedaque:2002mn}. In this
theory nucleons are described as point particles with zero-range
interactions whose strengths are adjusted to reproduce the scattering
lengths $a_{t}$ and $a_{s}$.  The effective ranges $r_s$, $r_t$ and
higher-order terms in the low-energy expansions of the phase shifts
are treated as perturbations.  Since the spin-singlet and spin-triplet
$np$ scattering lengths of $a_{s} = -23.8$ fm and $a_{t} = 5.4$ fm are
significantly larger than the $NN$-interaction's range $R\approx
2$~fm, \eftnopinew's expansion in $R/|a|$ is a useful tool for analyzing
the $NN$ system at energies $\ll 1/(M R^2)$.  In fact, the utility of
this EFT is not confined to nuclear physics, since systems where
$R/|a|$ is a small parameter also occur in atomic, molecular, and
particle physics. In the limit that $R/|a| \rightarrow 0$ such systems
exhibit ``universal'' features, which are completely independent of
details of the inter-particle potential~\cite{Braaten:2004rn}.  The
calculations of Refs.~\cite{Beane:2001bc,Beane:2002xf,Epelbaum:2002gb}
suggested that $NN$ systems with a common light-quark mass $m_u=m_d$
chosen to yield $m_\pi \simeq 200$ MeV would be close to this
universality limit.

A particularly striking feature of this limit occurs in the
three-nucleon system.  In the 1970s Efimov showed that if $|a|$ is
much larger than the range $R$ of the interaction there are shallow
three-body bound states whose number increases logarithmically with
$|a|/R$.  In the `resonant limit' $a \to \pm \infty$, there are
infinitely many shallow three-body bound states with an accumulation
point at the three-body scattering threshold.  If the particles are
identical bosons, the ratio of the binding energies of successive
states rapidly approaches the universal constant $\lambda_0^2 \approx
515$.  Efimov also showed that low-energy three-body observables for
different values of $a$ are related by a discrete scaling
transformation in which $ a \to \lambda_0^n a$, where $n$ is an
integer, and lengths and energies are scaled by the appropriate powers
of $\lambda_0$ \cite{Efi71,Efi79}.  The discrete scaling symmetry is
the hallmark of an RG limit cycle \cite{Wilson:1970ag}.  

The Efimov effect can also occur for fermions with at least three
distinct spin or isospin states and therefore applies to nucleons as
well. The extension of \eftnopi to the three-nucleon system
allows a systematization of Efimov's ``qualitative'' approach to the
three-nucleon problem \cite{Efi81}. 
In \eftnopinew, treatment
of the $S_{1/2}$ $nd$ partial-wave requires the inclusion of a
three-body force at leading order in the power counting
\cite{Bedaque:1999ve}. It is this three-body force (commonly denoted
$H_0$), whose renormalization-group evolution is governed by a limit
cycle. Thus, once one piece of three-body data is used to fix the
three-body force at a given regularization scale, other three-body
observables can be predicted in \eftnopinew. This provides an
explanation for correlations observed empirically in the three-nucleon
system, e.g. the ``Phillips line'' correlation between $B_3$ and the
doublet $nd$ scattering length $a^{1/2}_{nd}$.

But the results such as the Phillips line which were discussed in
Ref.~\cite{Bedaque:1999ve} were only obtained there up to corrections
of order $R/|a|$ and $k R$, where $k$ is the typical momentum of the
process under consideration.   The corrections to
three-nucleon scattering observables linear in $R/|a|$ were
considered in Refs.~\cite{EfT85,Hammer:2001gh}. Efforts to compute
$(R/|a|)^2$ and $(k R)^2$-corrections to three-nucleon observables
were pursued in Ref.~\cite{Bedaque:2002yg}. However, recent work
within a reformulation of the equations describing three-nucleon
scattering shows that, once one three-body datum is used to fix the
leading-order three-nucleon force, the inclusion of corrections up to and
including $(R/|a|)^2$ effects is straightforward and does not require
any additional three-body input~\cite{Platter:2006a,Platter:2006b}.
(For a general analysis of the order in
$R/|a|$ at which three-body input is first needed in a given $nd$
partial wave, see Ref.~\cite{Griesshammer:2005}.)  

These efforts open the way for precision calculations of three-body
observables in \eftnopi.  Calculations of three-body
observables in this EFT are much simpler than in the $\chi$EFT, and
the computational effort is significantly smaller. This is
particularly so near the critical trajectory, where the $\chi$EFT is
trying to bridge a huge range of distance scales: from a
short-distance scale $1/\Lambda_\chi$ of fractions of a fermi to an
$NN$ scattering length of literally hundreds of fermis. The pionless
EFT provides a natural way to build in the physics that occurs between
the distance scale $R$ and the unnaturally large scattering
lengths that occur in this regime.

Therefore the $\chi$EFT and \eftnopi offer mutually
complementary approaches to an understanding of the quark-mass dependence of
few-nucleon system observables.  In order to use \eftnopi to
understand quark-mass dependence we, of course, need input from
the $\chi$EFT, but once that input is in hand predictions for
a variety of few-nucleon-system observables can be derived up to a
given order in the $R/|a|$ expansion.

A first exploratory study along these lines was carried out in in
\eftnopi in Ref.~\cite{Braaten:2003eu}.  The quark-mass
dependence of the nucleon-nucleon scattering lengths from
Ref.~\cite{Epelbaum:2002gb} was used as input to that calculation.  In
Ref.~\cite{Epelbaum:2006jc}, a detailed investigation of the
possibility that $1/a$ could vanish in both the $\trip$ and
$\sing$ channels, {\it at the same $m_\pi$} was performed using an
$NN$ potential computed up to next-to-leading order (NLO) in the
chiral expansion. The ``critical'' pion mass studied most thoroughly
in Ref.~\cite{Epelbaum:2006jc} was $\mpic=197.8577$ MeV.  For that
case the NLO $\chi$EFT calculation was matched to a leading-order (LO) \eftnopi
computation. The $m_\pi$-dependence of $1/a$ in both channels,
as well as the pion-mass dependence of the triton binding energy, was
used as input to that computation, and the spectrum
of excited three-body states for $m_\pi$ close to $\mpic$
was computed. At LO that spectrum consists of an infinite tower
of states with binding energies that obey:
\begin{equation}
\frac{B_3^{(n)}}{B_3^{(n+1)}} \approx 515.035,
\label{eq:efimovratio}
\end{equation}
where $B_3^{(n)}$ is the binding energy of the $n$th excited state in
the tower. This LO prediction will be modified by higher-order terms
in \eftnopi which are larger for the more deeply
bound states. Examining the `data' of Ref.~\cite{Epelbaum:2006jc} for
this ratio we see that it is 543 for $n=0$ and $5.2 \times 10^2$ for
$n=1$. Here we show that \eftnopi computations of the ratio
$B_3^{(n)}/B_3^{(n+1)}$ converge rapidly for all $n \neq 0$, with the
prediction (\ref{eq:efimovratio}) receiving corrections of at most
0.1\% at NLO, and vanishingly small corrections at orders beyond that.

This extends the work of Ref.~\cite{Epelbaum:2006jc}, because we use
the same NLO $\chi$EFT calculation as input, but compute observables
up to next-to-next-to-leading order (N$^2$LO) in \eftnopinew.  The work of
Refs.~\cite{Platter:2006a,Platter:2006b} shows that in order to do this
the only additional information we need is the effective range in the
$\trip$ and $\sing$ channels as a function of $m_\pi$. With these two
effective ranges, together with knowledge of $1/a_{s}(m_\pi)$ and
$B_d(m_\pi)$ and $B_3(m_\pi)$, we can use \eftnopi
to predict the behavior of the excited states of the three-body system
and the doublet $nd$ scattering length $a^{1/2}_{nd}$ as functions of
$m_\pi$ in a domain of $m_\pi$ around $\mpic$. We also provide NLO
predictions for the quartet $nd$ scattering length $a^{3/2}_{nd}$ as
a function of $m_\pi$.

In the process of providing these predictions we are presented with
the opportunity to examine the convergence pattern of \eftnopi
in the three-nucleon sector. Since consideration of different
pion masses yields different values of $R/|a|$, \eftnopi
will not converge at the same rate at all values of the pion mass. Furthermore,
the presence of multiple three-nucleon bound states near the QCD critical
trajectory provides us with the opportunity to examine how this EFT
converges for bound states with different binding energies. In fact,
we will find that near the critical trajectory the limiting factor in the
accuracy of the calculations comes from corrections which scale as
$\kappa R$ (with $\kappa$ the momentum characteristic of the bound
state being studied) and not from the often-quoted expansion parameter
$R/|a|$, which is actually very small for $m_\pi$ near $\mpic$.

This paper is structured as follows. In
Section~\ref{sec-chiral} we review the pertinent results obtained in
chiral effective field theories for quark-mass dependence of
observables in systems with $A=0,1,2$, and $3$. In
Sec.~\ref{sec-nopi}, we explain how these results are used as input to
equations that describe neutron-deuteron scattering (and
neutron-deuteron bound states) up to N$^2$LO in \eftnopi.  We present
the results of our N$^2$LO computation in Sec.~\ref{sec-results},
together with a discussion of their convergence pattern and conclude
in Sec.~\ref{sec-conclusion}.

\section{Chiral Effective Field Theory}

\label{sec-chiral}

In order to understand the quark-mass dependence of hadronic
observables we need a theory that encodes the soft breaking of QCD's
$SU(2)_L \times SU(2)_R$ chiral symmetry by the quark mass term in the
QCD Lagrangian. Chiral effective field theory is a low-energy theory
with the same (low-energy) symmetries and pattern of symmetry breaking
as QCD. As such it provides a systematic way to compute the quark-mass
dependence of observables in few-nucleon systems. 

In the $A=0$ and $A=1$ sectors the technology to do this is
well-established, coming under the name ``chiral perturbation
theory''~\cite{Bernard:2006gx}.
Scattering amplitudes are expanded in powers of the small parameter:
\begin{equation}
P \equiv \frac{m_\pi,p}{4 \pi f_\pi, m_\rho}
\end{equation}
where $p$ is the typical momentum of the process under consideration.
Since, according to the Gell-Mann-Oakes-Renner relation,
$m_\pi^2 \sim m_q$, chiral
perturbation theory provides access to quark-mass dependence, up to
an accuracy that is determined by the order to which the computation
is carried out. 

In the $A=0$ sector many observables have now been computed up to 
two loops, which is equivalent to a computation up to effects of
$O(m_\pi^6) \equiv O(m_q^3)$. In the case of $A=1$ most computations 
have been performed up to ``complete one-loop order'', which means
that they include all effects up to $O(m_\pi^4) \equiv O(m_q^2)$. 

For $A \geq 2$ the presence of non-perturbative effects makes the
chiral counting more interesting. Weinberg proposed that the chiral
perturbation theory expansion could be applied to the nucleon-nucleon
potential in the $A=2$ system, and, more generally, to the overall
nuclear potential in a many-body system~\cite{Weinberg:1991um}. In
this counting three-nucleon forces do not occur until
next-to-next-to-leading order in the chiral expansion, and are
suppressed by three powers of the small parameter~\cite{vanKolck:1994yi}. 
The calculations of
Refs.~\cite{Braaten:2003eu,Epelbaum:2006jc} were based on a nuclear
force computed up to $O(m_\pi^2) \equiv O(m_q)$ in the chiral
expansion~\cite{EGM2}.

In this section, we review the relevant results on the quark-mass
dependence of hadronic observables that entered that
calculation. Since the results given there for the $NN$ and $NNN$
systems are only valid up to effects of $O(m_q)$ we in general do not
provide expressions that go beyond this accuracy. We will see that
present uncertainties in the quark-mass dependence of observables in
the $A=2$ system preclude reliable predictions for the
$m_q$-dependence of three-body binding energies and scattering
lengths.  However, as long as there is a domain of $m_q$ values where
both $NN$ scattering lengths become large with respect to $1/m_\pi$
then the approach of the Section~\ref{sec-nopi}, that uses \eftnopinew, 
together with a small amount of input from $\chi$EFT, to
predict the quark-mass dependence of a variety of $A=3$ observables,
will always be applicable. This is true irrespective of the precise
value of $\mpic$---or even whether the critical trajectory is only
approximately realized. Therefore, \eftnopi can always be used to
propagate information from the $\chi$EFT through to few-nucleon system
observables---as long as $|a| m_\pi \gg 1$ and $p/ m_\pi \ll 1$.
And, as the accuracy of
the $\chi$EFT input to \eftnopi improves, the results
obtained using the expansion around the universal limit $|a|=\infty$
will become a better approximation to the quark-mass dependence of
full QCD.

\paragraph{$A=0$} 
At the order we work to here, quark-mass dependence is synonymous
with pion-mass dependence because of the Gell-Mann-Oakes-Renner
relation: 
\beq 
m_\pi^2 = -(m_u + m_d) \langle 0 | \bar{u} u | 0
\rangle/f_\pi^2\,,
\label{eq:mq-mpi}
\eeq where $\langle 0 | \bar{u} u | 0 \rangle \approx (-225 \mbox{
  MeV})^3$ is the quark condensate, and $f_\pi=92.4$ MeV is the
pion-decay constant. In the following, we will therefore discuss all
our results in terms of pion-mass dependence. This is more convenient
for nuclear applications, since the pion mass is easy to adjust. The
corresponding change in quark masses can then be read off from
Eq.~(\ref{eq:mq-mpi}).

\paragraph{$A=1$}
The nucleon mass' dependence on $m_\pi$ has now been calculated to complete
two-loop order~\cite{Schindler:2006ha}. However,
here we need only the result up to $O(m_q)$. This can be written as:
\begin{equation}
M=M_0 - 4 c_1 m_\pi^2 + O(m_\pi^3),
\label{eq:Mnucpidep}
\end{equation}
where the leading non-analytic piece at $O(m_\pi^3)$ is easily
calculated from the one-loop nucleon self-energy (in dimensional
regularization with minimal subtraction), and the LEC $c_1$ is related
to the nucleon $\sigma$-term. Meanwhile, $M_0$ is the nucleon mass in
the chiral limit. If we take $c_1=-0.81~{\rm
  GeV}^{-1}$~\cite{Buettiker:1999ap}, we have $M_0=880$ MeV.
Correspondingly, $M(m_\pi)$ increases to over a GeV at the putative
critical pion mass $m_\pi \simeq 200$~MeV. Epelbaum {\it et al.}
claim that this has only a small effect on spectra and so do not
consider the effect of this change of
$M$~\cite{Epelbaum:2006jc,Epelbaum:2002gb}. Below we will
provide an independent assessment of how this change in $M$
affects the results near $m_\pi=\mpic$.

The other $A=1$ quantity of relevance for our study of nuclear bound states
is the pion-nucleon-nucleon coupling. For this we have the result:
\begin{equation}
\frac{g_{\pi NN}}{M}=\frac{g_A}{f_\pi} \left(1 + 2 \Delta - \frac{2
  m_\pi^2}{g_A} \bar{d}_{18}\right),
\label{eq:GT}
\end{equation}
where $g_A=1.26$ is the physical value of the nucleon's axial
coupling.  This is the Goldberger-Treiman relation in the chiral
limit, and here it is supplemented by an $O(m_q)$ term involving the
LEC $\bar{d}_{18}$. Using a value of $g_{\pi N N}$ extracted from a
phase-shift analysis of low-energy $\pi$N data~\cite{Matsinos:1998wp,Arndt:2006bf}
we get $\bar{d}_{18}=-0.97~{\rm GeV}^{-2}$~\cite{Epelbaum:2002gb}.
Also appearing in Eq.~(\ref{eq:GT}) is the ratio $\Delta$, which
encodes the $m_q$-dependence of the ratio $g_A/f_\pi$. $\Delta$ is
just the fractional change in this ratio at an arbitrary value of
$m_\pi$, as compared to the value at the physical $m_\pi$,
$m_\pi^{phys}$, i.e.:
\begin{equation}
\Delta \equiv \frac{(g_A/f_\pi)_{m_\pi} - 
  (g_A/f_\pi)_{m_\pi^{phys}}}{(g_A/f_\pi)_{m_\pi^{phys}}}
\end{equation}
The expression for $\Delta$ in terms of low-energy constants 
appearing in ${\cal
  L}_{\pi N}^{(3)}$ and ${\cal L}_{\pi \pi}^{(4)}$ can be found in
Ref.~\cite{Epelbaum:2006jc}.

\paragraph{$A=2$}
The quark-mass dependence of the chiral $NN$ interaction was
calculated to next-to-leading order (NLO) in the chiral counting in
Refs.~\cite{Beane:2001bc,Beane:2002xf,Epelbaum:2002gb}. In addition to
the quark-mass dependence of the pion mass and $\pi NN$ coupling that
appear in the one-pion-exchange potential the short-distance part of
the force acquires a quark-mass dependence of its own. This is
essential for correct renormalization~\cite{Kaplan:1996xu}. This
quark-mass dependence can be described by parameters $D_{s}$ and
$D_{t}$, so that the $NN$ potential in a particular S-wave 
($t$=$\trip$, $s$=$\sing$) is:
\begin{equation}
V=C_{s/t} + D_{s/t} \left[m_\pi^2 -
  (m_\pi^{phys})^2\right] + V_{OPE} + V_{TPE}^{(2)},
\end{equation}
where $V_{OPE}$ is the (partial-wave projected) one-pion exchange with
the already discussed variation in $g_{\pi NN}$ (\ref{eq:GT}) and
$m_\pi$ (\ref{eq:mq-mpi}) taken into account, and $V_{TPE}^{(2)}$ is
the well-known ``leading'' two-pion
exchange~\cite{Ordonez:1995rz,Kaiser:1997mw,EGM2}.  Note that to the
order considered here it is only necessary to take into account the
$m_\pi$-dependence of $g_{\pi NN}$ (or, equivalently, $g_A$ and $f_\pi$)
in the LO piece of the $NN$
potential.  We do not take into account the $m_\pi$-dependence of
$g_A$ and $f_\pi$ in the two-pion-exchange potential
but instead simply use their physical values there. Considering only
``explicit'' $m_\pi$ dependence in $V_{TPE}^{(2)}$ is sufficient
at the order we work to here.

Meanwhile, in
Ref.~\cite{Epelbaum:2006jc} the unknown parameters $D_{s/t}$
were converted into dimensionless numbers via:
\begin{equation}
\alpha_{s/t} \equiv \frac{f_\pi^2 \Lambda_\chi^2
  D_{s/t}}{16 \pi}
\end{equation}
At present next-to-nothing is known about the values of $\alpha$. But
predictions for the pion-mass dependence of S-wave scattering can be
obtained by varying the $\alpha$'s within naturalness
bounds. The uncertainty in the behavior of $\Delta$ as a function of
$m_\pi$ then further enlarges the range of results obtained for, say,
the $NN$ scattering lengths as a function of $m_\pi$.

Because of these uncertainties, the study of
Ref.~\cite{Epelbaum:2006jc} did not attempt to predict the precise
value of the critical pion mass where the inverse scattering
lengths in both S-waves vanish. This strategy is motivated not just by
ignorance as regards the two LECs appearing in the $\alpha$'s, but
also because we have not considered the impact that $m_u \neq m_d$
would have on the $NN$ potential. Such effects must be included if the
Braaten-Hammer conjecture, as originally formulated, is to be
investigated~\cite{Braaten:2003eu}. Instead, in
Ref.~\cite{Epelbaum:2006jc} $\alpha$ parameters were chosen in such a
way that a critical trajectory occurred.  For critical pion masses in
the range 175 MeV $\leq \mpic\leq$ 205 MeV, it is always possible to
find an appropriate set of $\alpha$ parameters.  Three sets of
$\alpha$ parameters were given in Ref.~\cite{Epelbaum:2006jc}, 
with each one corresponding to a
different critical pion mass $\mpic$.  The computations of
Ref.~\cite{Epelbaum:2006jc} focused particularly on a 
choice $\alpha_{t}=-2.5$, which yields $\alpha_{s}=2.138598$
and a critical pion mass $\mpic = 197.8577$ MeV.
Although it is unlikely that physical QCD will correspond to this
particular solution, many aspects of the limit cycle are universal and
do not depend on the exact parameter values
\cite{Braaten:2004rn}. Therefore in what follows we take the results
reported in Ref.~\cite{Epelbaum:2006jc} for this choice of $\chi$EFT
parameters as both the input data for \eftnopi and the experimental
data to which we will compare our \eftnopi results for the pion-mass
dependence of three-nucleon observables in the vicinity of the QCD
critical trajectory.

As already discussed in the introduction, the $NN$ inputs we need from
the $\chi$EFT for our NNLO \eftnopi calculation are scattering lengths
and effective ranges in the spin-triplet and spin-singlet
channels. The scattering lengths $a_t$ and $a_s$ can, of course, be
exchanged for the momenta $\gamma_t$ and $\gamma_s$ characterizing the
position of the bound-state/virtual-state poles in the two-body
propagator. If higher-order terms in the effective-range expansion are
neglected, the relation between the two quantities is simply \beq
\gamma_{\alpha}=\frac{1}{r_{\alpha}}\left(1-\sqrt{1-2r_{\alpha}/a_{\alpha}}\,\right)\,,
\eeq with $\alpha=s,t$ indicating a particular $NN$ channel.  The
advantage of using $\gamma_{\alpha}$, rather than $a_\alpha$, is that
the $NN$ pole position is not modified by higher-order corrections.
In the following, we will therefore use $\gamma_t(m_\pi)$,
$\gamma_s(m_\pi)$, $r_t(m_\pi)$, and $r_s(m_\pi)$ as the $NN$ input.

\begin{figure}[tb]
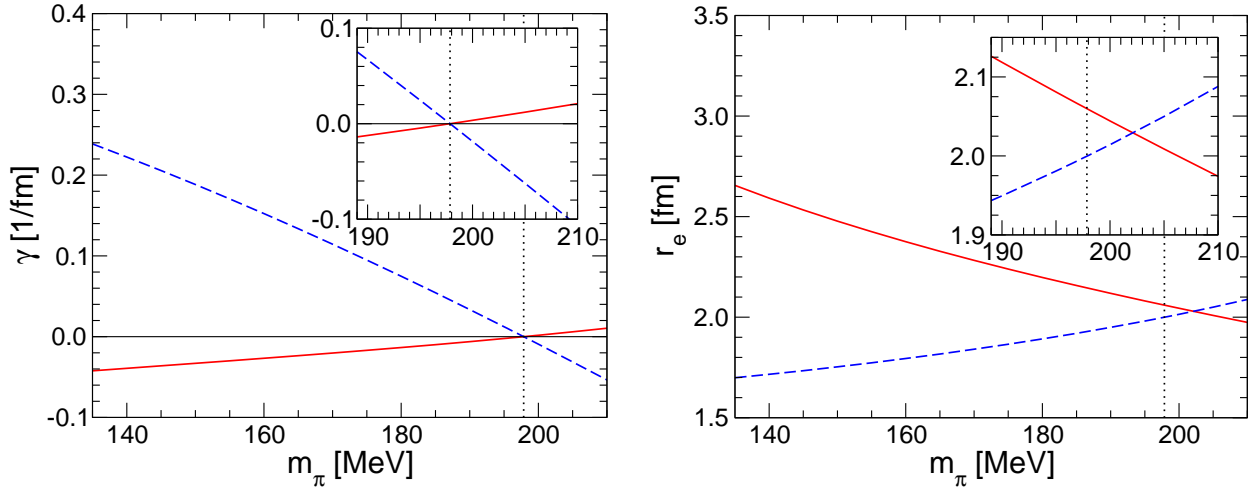

\centerline{
\includegraphics*[width=8cm,angle=0,clip=true]{gamsgamt_mpi_full.eps}
\quad
\includegraphics*[width=8cm,angle=0,clip=true]{resret_mpi_full.eps}
}
\caption{Left panel: Pole momenta in the 
$\trip$ (dashed line) and $\sing$ (solid line)
nucleon-nucleon channels as a function of the pion mass $m_\pi$ 
(from Ref.~\cite{Epelbaum:2006jc}).
The vertical dotted line indicates the critical pion mass $\mpic$
while the inset shows the critical region in more detail.
Right panel: same for the effective ranges
(from Ref.~\cite{EvgeniPrivate}).
}
\label{fig:gammasgammat}
\end{figure}

The behavior of the input pole momenta $\gamma_s(m_\pi)$ and
$\gamma_t(m_\pi)$ in the vicinity of 
$m_\pi=\mpic$ is depicted in the inset in the left panel of 
Fig.~\ref{fig:gammasgammat}.
As promised, the pole momenta vanish at the critical value of the pion
mass $\mpic = 197.8577$ MeV.  For pion masses below $\mpic$
$\gamma_t(m_\pi) > 0$ and the deuteron is 
bound.  As $\gamma_t(m_\pi)$ decreases, the deuteron becomes
more and more shallow and finally becomes unbound at the critical pion
mass. Above the critical pion mass the deuteron exists as a shallow
virtual state.  In the spin-singlet channel, the situation is
reversed: the \lq\lq spin-singlet deuteron'' is a virtual state below
the critical pion mass and becomes bound above.  
In the right panel of
Fig.~\ref{fig:gammasgammat} we show the corresponding dependence of
the spin-singlet and spin-triplet effective ranges
\cite{EvgeniPrivate}.  The pion-mass dependence of the effective
ranges is weak and both are close to 2~fm throughout the region
displayed. In the region of interest $r_s(m_\pi)$ decreases
monotonically as the pion mass is increased whereas $r_t(m_\pi)$
increases.  At a pion mass of about 202 MeV the curves
cross. 

Above we pointed out that the results of
Refs.~\cite{Epelbaum:2006jc,EvgeniPrivate} which are presented in
Fig.~\ref{fig:gammasgammat} do not include effects due to the
pion-mass dependence of the nucleon mass.  Such effects can easily be
accounted for though, by noting that the computations presented in
Fig.~\ref{fig:gammasgammat} are done in a non-relativistic framework,
and no $1/M$ corrections to the $NN$ potential are
incorporated. Consequently the nucleon mass can be scaled out of the
problem by an appropriate choice of units. The critical pion mass is
then written as:
\begin{equation}
\mpic=0.2107 M,
\label{eq:mpic}
\end{equation}
where we have employed an (isospin-averaged) nucleon mass of $M=938.9$
MeV.  A value of $\mpic$ in MeV that includes the impact of $M(m_\pi)$
can be obtained from Eq.~(\ref{eq:mpic}) by inserting the expression
(\ref{eq:Mnucpidep}) for $M(\mpic)$ in place of $M$ on the right-hand
side, and solving the resulting equation for $\mpic$. This yields
$\mpic=220$ MeV, with the accuracy limited to two
digits by the precision with which $c_1$ is known.

We can use the same argument to assess the impact of the
$m_\pi$-dependence of $M$ on two-body observables when $m_\pi \neq
\mpic$. Incorporating the pion-mass dependence of the nucleon mass in
the calculation of Refs.~\cite{Epelbaum:2006jc,EvgeniPrivate} in this
way would thus lead to a change of the scale on both axes in the left
panel of Fig.~\ref{fig:gammasgammat} by a factor of $M(m_\pi)/M$.  The
x-axis in the right panel would be rescaled by the same factor, while
the y-axis there would be rescaled by $M/M(m_\pi)$.  The effect of
such rescaling is largest at the highest pion masses considered, but
even there it is less than 15\%. Other uncertainties in the
calculation of the functions $\gamma_s(m_\pi)$, $\gamma_t(m_\pi)$,
$r_s(m_\pi)$, and $r_t(m_\pi)$ are larger than this, and so from here
on we follow the procedure of
Refs.~\cite{Epelbaum:2006jc,EvgeniPrivate} and ignore effects on these
quantities due to the pion-mass dependence of $M$.

\begin{figure}[tb]
\centerline{\includegraphics*[width=10cm,angle=0,clip=true]{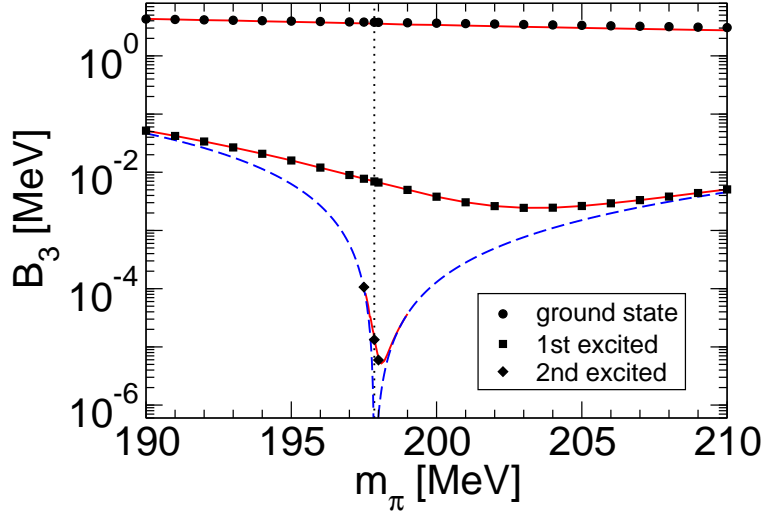}}
\caption{Triton ground and excited state energies $B_3$ in the
  critical region. The circles (ground state), squares (first excited
  state), and diamonds (second excited state) give the $\chi$EFT
  result, while the solid lines are calculations in the pionless
  theory to N$^2$LO.  The vertical dotted line indicates the critical
  pion mass $\mpic$.  The thresholds for three-nucleon states to be
  stable against breakup into a single nucleon plus an $NN$ bound
  state are given by the dashed lines. The zero of energy corresponds
  to the opening of the 3N channel.}
\label{fig:bind3nnlo}
\end{figure}

\paragraph{A=3}
We also need one three-body datum as input to \eftnopi in order to
perform our N$^2$LO computation of three-body system observables.  In
Ref.~\cite{Epelbaum:2006jc}, the binding energies of the triton and
the first two excited states in the vicinity of the limit cycle were
obtained from the solution of the Faddeev equations for the NLO
$\chi$EFT potential.  For much of the pion-mass range of interest to
us there is only one bound state, and so, in that region, $100 \leq
m_\pi \leq 190$ MeV, we renormalize \eftnopi using the binding energy
of the triton ground state, $B_3^{(0)}$, and then predict the $nd$
doublet scattering length, $a_{nd}^{1/2}$. But our results for
pion-mass dependence are particularly interesting in the ``critical region'',
which we define to be values of $m_\pi$ for which the triton has at
least one excited state. For the choice $\mpic=197.8577$ MeV the
calculations of Ref.~\cite{Epelbaum:2006jc} indicate a critical region
corresponding to $190~{\rm MeV} \leq m_\pi \leq 210~{\rm MeV}$.  The
first excited state that is present in this region is much shallower
than the ground state, and is transparently within the domain of
validity of \eftnopi.  Thus errors in that EFT are smaller if we
renormalize to its binding energy, $B_3^{(1)}$, rather than to
$B_3^{(0)}$.  The \eftnopi result for $B_3^{(0)}$ then becomes a
prediction and we can also predict $a_{nd}^{1/2}$.  The
$\chi$EFT results for both the ground and first-excited state in the
``critical region'' are shown in Fig.~\ref{fig:bind3nnlo}, together
with our N$^2$LO \eftnopi calculation.


\section{The Pionless EFT to N$^2$LO}
\label{sec-nopi}
In the critical region where there is more than one three-nucleon
bound state, the length scales in the three-nucleon problem change
rapidly as $m_\pi$ is varied.  The binding energy of the deepest lying
three-nucleon state changes slowly, and is still within a factor of
two of its experimental value when $m_\pi^{crit}$ is reached. However,
in this region the
neutron-deuteron scattering length varies between $-\infty$
and $\infty$, and does so an infinite number
of times. A single cycle between -$\infty$ and
$\infty$ takes place over the range of pion-mass increase that is needed to
make an additional shallow three-nucleon (``Efimov'') state appear as
$m_\pi \rightarrow \mpic$.  Consequently, the numerical effort which
has to be devoted to the calculation of three-nucleon observables
increases significantly once $m_\pi$ is large enough that the first
Efimov state appears---or is close to appearing. We advocate
determining the dependence of key two- and three-nucleon observables
on the pion mass with $\chi$EFT, then using those results as input
to the simpler \eftnopi and employing that EFT to analyze the behavior
of other three-nucleon observables in the critical region.

\eftnopi is a systematic expansion in contact interactions where---for
momenta $k \sim 1/a$---the small expansion parameter is $R/a$.
The corresponding low-energy Lagrangian for the
neutron-deuteron system is given by~\cite{Bedaque:1999ve} \bea
\nonumber \mathcal{L}&=&N^\dagger(i\partial_0+\frac{\nabla^2}{2M})N
-t_i^\dagger(i\partial_0+\frac{\nabla^2}{4M}-\Delta_t)t_i
-s_j^\dagger(i\partial_0+\frac{\nabla^2}{4M}-\Delta_s)s_j\\ \nonumber
&& +g_t \left( t_i^\dagger N^T \tau_2 \sigma_i \sigma_2
N+\text{h.c}\right) +g_s \left( s_j^\dagger N^T \sigma_2 \sigma_j
\tau_2 N+\text{h.c}\right) \\ &&-G_3
N^\dagger\left(g_t^2(t_i\sigma_i)^\dagger t_{i'}\sigma_{i'}
+\frac{1}{3}g_t g_s[(t_i\sigma_i)^\dagger s_j\tau_j+\text{h.c}]
+g_s^2(s_j\tau_j)^\dagger s_{j'}\tau_{j'}\right)N+\ldots~,
\label{lagrange}
\eea
where $N$ represents the nucleon field and $t_i (s_j)$ are the di-nucleon
fields for the $\trip$ ($\sing$) channels with the corresponding
quantum numbers, respectively. The $\sigma_i$ ($\tau_j$) are Pauli
matrices in spin (isospin) space, respectively, and the
dots indicate additional terms with more
fields/derivatives. 

To renormalize three-nucleon observables in the $nd$ S$_{1/2}$ channel within
this framework, a three-nucleon force symmetric under $SU(4)$
spin-isospin rotations \cite{Bedaque:1999ve} (here denoted by $G_3$)
or {\it equivalently} a subtraction has to be
performed~\cite{Hammer:2000nf,Afnan:2003bs}.  Therefore, one
three-body input parameter is needed for the calculation of
observables in this channel---the one in which the triton exists.

In order to compute three-nucleon observables we need the full
two-body propagator $\tau$, which is the result of dressing the bare
di-nucleon propagator by nucleon loops to all orders. The EFT is
arranged to reproduce the effective-range expansion for $\tau$, i.e.
\beq \tau_{\alpha}(E)=-\frac{2}{\pi
  M}\,\frac{1}{-\gamma_\alpha+\sqrt{-ME}
  +\frac{r_\alpha}{2}\left(\gamma_\alpha^2+ME\right)}~,
\label{eq:TBamp}
\eeq where $E$ denotes the two-body energy in the two-body c.m. frame,
$\gamma_\alpha$ gives the $NN$ pole position, $r_\alpha$ is the effective range,
and the index $\alpha=s,t$ indicates either the singlet or triplet
$NN$ channel.  In the triplet channel the pole position
$\gamma_t$ is related to the deuteron binding energy
$B_d=\gamma_t^2/M$.  The form (\ref{eq:TBamp}) cannot be directly used
as input to EFT three-nucleon calculations because it has poles at
energies outside the domain of validity of the EFT. Therefore, the
propagator cannot be employed within a three-body integral equation which
has cutoffs $\Lambda > 1/r_\alpha$ unless additional techniques to
subtract these unphysical poles are implemented. Instead of using the
propagator in the form above we will expand it to $n$th order in $R/a$:
\beq
\label{eq:tau}
\tau^{(n)}_\alpha(E)=\frac{S^{(n)}_\alpha(E)}{E+\gamma_\alpha^2/M}.
\eeq
For $n<3$, the residue $S^{(n)}$ is given by
\beq
S^{(n)}_\alpha(E)=\frac{2}{\pi M^2} \sum_{i=0}^n \left(\frac{r_\alpha}{2}\right)^i
[\gamma_\alpha + \sqrt{-ME}]^{i+1}~,
\label{eq:Sn}
\eeq
while for $n\geq 3$ higher-order terms in the effective-range expansion
contribute to $S(E)$.
The set of integral equations for the nucleon-deuteron K-matrix 
generated by this EFT (neglecting, for the moment, the $nd$ coupling $G_3$)
is given by \cite{Bedaque:1999ve,Afnan:2003bs}
\bea
\nonumber\label{eq:integ}
K_{tt}^{(n)}(q,q';E)&=&\mathcal{Z}_{tt}(q,q';E)
+\mathcal{P}\int_0^\Lambda\hbox{d}q''\,q''^2
\mathcal{Z}_{tt}(q,q'';E)
\tau_t^{(n)}(E-\frac{3}{4}\frac{q''^2}{M})K_{tt}^{(n)}(q'',q';E)\\
\nonumber
&&\quad\qquad\qquad+\mathcal{P}\int_0^\Lambda\hbox{d}q''\,q''^2
\mathcal{Z}_{ts}(q,q'';E)
\tau_s^{(n)}(E-\frac{3}{4}\frac{q''^2}{M})K_{st}^{(n)}(q'',q';E)~,
\eea
\bea
\label{eq:integraleq}
\nonumber
K_{st}^{(n)}(q,q';E)&=&\mathcal{Z}_{st}(q,q';E)
+\mathcal{P}\int_0^\Lambda\hbox{d}q''\,q''^2\mathcal{Z}_{st}(q,q'';E)
\tau_t^{(n)}(E-\frac{3}{4}\frac{q''^2}{M})K_{tt}^{(n)}(q'',q';E)\\
\nonumber
&&\quad\qquad\qquad+\mathcal{P}\int_0^\Lambda\hbox{d}q''\,q''^2
\mathcal{Z}_{ss}(q,q'';E)
\tau_s^{(n)}(E-\frac{3}{4}\frac{q''^2}{M})K_{st}^{(n)}(q'',q';E)~,
\nonumber\\
\eea
where $n$ is the order of the calculation (in what follows we assume
$n<3$)
and $\mathcal{P}$ indicates a principal-value integral.
The formulation in terms of the K-matrix is useful as long as we are
only interested in bound states or $nd$ scattering below the
three-nucleon threshold.
The Born amplitude $\mathcal{Z}_{\alpha\beta}$ is given by
\beq
\mathcal{Z}_{\alpha\beta}(q,q';E)=-\lambda_{\alpha\beta}\frac{M}{q q'}
\log\left(\frac{q^2+q q'+q'^2-ME}{q^2-q q'+q'^2-ME}\right)~,
\label{eq:Zalphabeta}
\eeq with the isospin matrix
\beq
\lambda_{\alpha=\{s,t\};\,\beta=\{s,t\}}=\frac{1}{4}\left(\begin{array}{cc} 1 &
  -3 \\ -3 & 1 \end{array}\right)~.  
\eeq 
Below three-nucleon breakup
threshold the K-matrix is related to $nd$ phaseshifts through 
\bea
K(k,k;E)&=&-\frac{3M}{8\gamma_t k}\tan\delta~.  
\eea 
In Refs.~\cite{Platter:2006a,Platter:2006b}, it was shown that the use
of one subtraction allows $\tau^{(n)}$ up to N$^2$LO ($n=2$) to
be inserted in Eqs.~(\ref{eq:integ}) with cutoff-independent
predictions resulting.  This is in marked contrast to the unsubtracted
equations written above, which generate cutoff variation of $O(1)$ in
low-energy observables if left
unrenormalized~\cite{Bedaque:1998kgkm,Bedaque:1999ve}.  The price that
is paid for the improved ultraviolet behavior of the subtracted
equations is that $a_{nd}^{1/2}$ appears as an input parameter in the
integral equation.  For further details on this subtraction method, we
refer to Refs.~\cite{Afnan:2003bs,Platter:2006a}.  In the next section
we will use the subtracted equations to perform calculations in the
pionless EFT up to N$^2$LO.

\begin{figure}[tb]
\centerline{\includegraphics*[width=12cm,angle=0,clip=true]{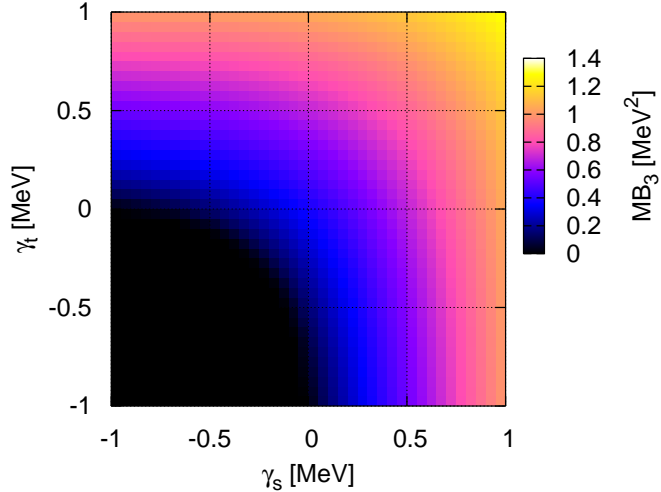}}
\caption{Binding energy $B_3$ of an excited 
state of the triton as a function of $\gamma_s$ and $\gamma_t$
in the vicinity of $\mpic$. The computation is performed here to
leading order in \eftnopi.}
\label{fig:su4symm}
\end{figure}

To close this section we note that the Lagrangian (\ref{lagrange}) is
symmetric under Wigner $SU(4)$ spin-isospin rotations provided
$g_s=g_t$ and $\Delta_s=\Delta_t$. Moreover, the integral equations
(\ref{eq:integraleq}) have an approximate $SU(4)$ symmetry for momenta
$q \gg \gamma_s,\gamma_t$, even if $g_s \neq g_t$ or $\Delta_s \neq
\Delta_t$.  (For a discussion of this symmetry for the $NN$ system see
Ref.~\cite{Mehen:1999qs}.)  Because of this symmetry, an
$SU(4)$-symmetric three-body force is sufficient for renormalization
\cite{Bedaque:1999ve}.  In Fig.~\ref{fig:su4symm}, we illustrate a
manifestation of this symmetry in the bound-state spectrum calculated
at LO in \eftnopinew. We show the binding energy of a triton excited
state in the critical region as a function of $\gamma_s$ and
$\gamma_t$.  In the lower left corner, the considered triton state
does not exist.  As $\gamma_s$ and $\gamma_t$ approach zero, the state
appears at threshold and then becomes more and more bound.  The figure
is clearly symmetric under reflection on the main diagonal
corresponding to a specific spin-isospin rotation that exchanges
$\gamma_s$ and $\gamma_t$.

Beyond leading order in \eftnopi $SU(4)$-breaking effects
enter the computation through the difference in singlet and triplet
effective ranges $r_s-r_t$. The right-hand panel of
Fig.~\ref{fig:gammasgammat} displays the interesting feature that---at
least for the choice of $\chi$EFT short-distance coefficients being
studied here---$r_s-r_t$ vanishes at a pion mass only a little larger
than $\mpic$. As remarked upon in Ref.~\cite{Epelbaum:2006jc}, this
results in $SU(4)$-breaking effects near the critical point that are
much smaller than one might naively expect, since $r_s - r_t \ll R$,
the range of the $NN$ interaction. More generally,
Fig.~\ref{fig:gammasgammat} suggests that a suitable choice of
$\alpha_s$ and $\alpha_t$ could lead to $r_s-r_t=\gamma_s=\gamma_t=0$ at
a single ``$SU(4)$ critical pion mass''. It is unclear whether a
$\chi$EFT that has this feature has anything to do with QCD. But the
\eftnopi valid near a critical trajectory which also has
$r_s=r_t$ could be built using $SU(4)$-symmetric operators for the
$R/a$ expansion, with $SU(4)$ breaking additionally suppressed by
another small parameter.
\section{Results}
\label{sec-results}
In this section, we solve the once-subtracted version of the integral
equations (\ref{eq:integraleq}) for pion masses in the range 100 MeV
$\leq m_\pi \leq$ 200 MeV.  Two-nucleon pole positions and effective
ranges for the relevant $\chi$EFT solution were displayed in
Fig.\ref{fig:gammasgammat}, and in the critical region we take as our
one three-body input the energy of the first triton excited state
$B_3^{(1)}$. We provide a detailed discussion of the critical
region, and also examine the behavior of \eftnopi for pion masses
between the critical region and the physical value.  In the latter case the
three-body datum chosen as input is the triton binding energy
$B_3^{(0)}$.

\subsection{Bound-State Spectrum}
Turning first to the critical region, we display the results
obtained from solving the homogeneous version of
Eq.~(\ref{eq:integraleq}) for negative energies. The spectrum of
triton states as a function of $m_\pi$ at N$^2$LO is shown in
Fig.~\ref{fig:bind3nnlo}, and compared to the $\chi$EFT result.

The binding energies from $\chi$EFT are given in
Fig.~\ref{fig:bind3nnlo} by the circles ($B_3^{(0)}$), squares
($B_3^{(1)}$), and diamonds ($B_3^{(2)}$) \cite{Epelbaum:2006jc}.  The
dashed lines indicate the neutron-deuteron ($m_\pi \leq \mpic$) and
neutron-spin-singlet-deuteron ($m_\pi \geq \mpic$) thresholds. For
$B_3$ less than these energies three-body states are unstable against
breakup into the $1 + 2$ configuration. Directly at the critical pion
mass, these thresholds coincide with the three-body threshold and the
triton has infinitely many excited states.  The solid lines show our
N$^2$LO calculation in the pionless EFT where the first triton excited
state was used as input.  Our results for the second excited state and
the ground state reproduce the $\chi$EFT results very well.  An
important point here is that the binding energy of the triton ground
state varies only weakly over the whole range of pion masses. Indeed
$B_3^{(0)}(\mpic) \approx 0.5 B_3^{(0)}(m_\pi^{phys})$. The excited states are
influenced by the $1+2$ threshold and their energies vary more strongly.

\begin{figure}[tb]
\centerline{\includegraphics*[width=10cm,angle=0,clip=true]{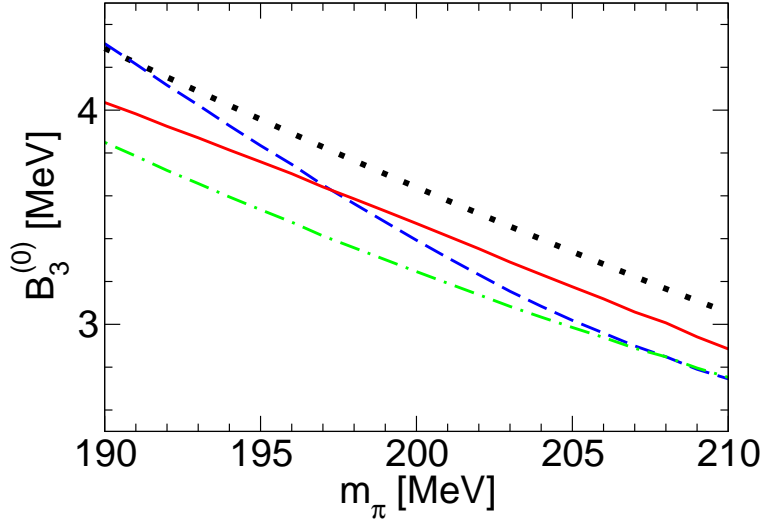}}
\caption{Binding energy of the triton ground state 
in the critical region computed in the
pionless EFT at LO (dashed), NLO (dash-dotted) and N$^2$LO (solid).
The dotted line gives the (interpolated) $\chi$EFT results
for comparison.}
\label{fig:pidep_bound}
\end{figure}

In order to study the convergence of the \eftnopi more thoroughly we
now look at the prediction for the triton ground-state energy in some
detail.  In Fig.~\ref{fig:pidep_bound}, we show our results for the
binding energy of the three-nucleon ground state for pion masses in
the ``critical region''
The dashed, dash-dotted, and solid lines denote, respectively, the LO,
NLO, and N$^2$LO results for the triton ground-state energy, 
while the dotted line gives the (interpolated) $\chi$EFT result of
Ref.~\cite{Epelbaum:2006jc}.  The leading-order \eftnopi
results describe $\chi$EFT's $B_3^{(0)}(m_\pi)$ curve
surprisingly well, although the shape of the $m_\pi$-dependence is
clearly wrong.  The NLO computation does not seem to yield significant
improvement in the overall agreement. The results of our N$^2$LO
calculation are on average better than the LO results. 

At first sight this slow convergence is surprising, since the
parameter $\gamma r \rightarrow 0$ at $m_\pi=\mpic$. But this simply
means that the slow convergence in the critical region must be due to
the significant (compared to $1/r$) binding momentum
$\kappa=\sqrt{4MB_3/3}$, which is about 65 MeV in this pion-mass range.

\subsection{Understanding Range Corrections in \eftnopi}
A rough estimate of the impact of higher-order corrections on \eftnopi
predictions can be obtained by applying naive dimensional
analysis. The relevant scales for the orders under consideration are
$\gamma$, $r$ and the binding momentum $\kappa$.  (For simplicity we
will drop the indices $s$ and $t$ in the ensuing discussion since the
scaling with $r_s$ and $r_t$ is the same.)  It follows therefore that
the error for an arbitrary three-body observable $O_3$ at order $j$
should scale as \beq \frac{\Delta O_3}{O_3}\approx A_{j+1}
\left(\frac{k r}{2}\right)^{j+1}+ B_{j+1} (\gamma r)^{j+1}\, ,
\label{err3}
\eeq
where $A_{j+1}$ and $B_{j+1}$ are (observable-dependent) numbers
of order one, and $k$ is the momentum associated with $O_3$.

However, this naive estimate clearly fails to explain the pattern of
convergence displayed in Fig.~\ref{fig:bind3nnlo}.
A more thorough understanding of this pattern of corrections can be gained by
examining the way that the NLO correction at the critical pion mass
scales with the parameters $\kappa$ and $r$. We will analyze a
perturbative expansion of \eftnopi about the LO
result~\cite{Hammer:2001gh}.  When $m_\pi=\mpic$, $\gamma=0$ and
$\gamma r$ corrections do not exist.  In this ``scaling limit'', the
two-body propagator at NLO simplifies to \beq
\tau^{(1)}(E)=-\frac{2}{\pi M}\left[\frac{1}{\sqrt{-M E}} +
  \frac{1}{\sqrt{-ME}} \left(-\frac{r}{2} ME\right)
  \frac{1}{\sqrt{-ME}}\right]~.
\label{eq:tau1crit}
\eeq
\begin{figure}[t]
\centerline{
\includegraphics*[width=150mm,clip=true]{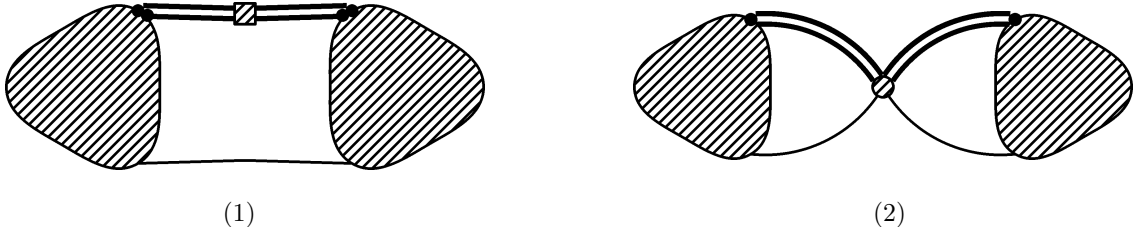}
}
\caption{
Diagrams contributing to the NLO shift in the binding energies.
The shaded square in Diagram 1 represents a non-trivial effective-range
insertion. The shaded circle in 
Diagram 2 denotes the insertion of the subleading three-body force
$H_1$. The shaded blobs are LO $nd$ vertex functions.
}
\label{fig:fmdiag}
\end{figure}

In Fig.~\ref{fig:fmdiag} we display the diagrams which have to be
taken into account for analysis of $\kappa r$ corrections.  Diagram
$(1)$ represents a perturbative insertion of $r/2$, \`{a} la the
second term of Eq.~(\ref{eq:tau1crit}).  Diagram $(2)$ denotes the
inclusion of the subleading three-body force $H_1$. The presence of
$H_1$ is necessary in order to absorb a divergence produced by diagram
$(1)$, but $H_1$ always appears in a fixed combination with $H_0$,
i.e. it does not need to be determined by any additional three-body
data~\cite{Hammer:2001gh}.

An analysis of the superficial degree of divergence of these diagrams
gives amplitudes which scale with momentum $p$ as:
\begin{equation}
{\cal M}^{(1)} \sim -\frac{r p}{2M}; \quad {\cal M}^{(2)} \sim
\frac{H_1 p^2}{M^2}.
\end{equation}
However, the only remaining momentum scale in the problem is the binding
momentum for the bound state under consideration, i.e. $\kappa^{(n)}$.
The NLO shift in the binding energy of the $n$th bound state is
therefore 
 \beq \Delta B_3^{(n)}=Z \left(\alpha
\frac{r}{2 M} \kappa^{(n)} - \beta \frac{H_1}{M^2} {\kappa^{(n)}}^2 \right)~, \eeq
where $\alpha$ and $\beta$ are numbers of order one, which, due to the
discrete scale invariance of the LO spectrum in \eftnopi,
are independent of $n$. 
$Z$ denotes
the wave-function renormalization, and a similar diagrammatic analysis
leads us to conclude that $Z \sim
\left(\kappa^{(n)}\right)^2$. We therefore obtain 
\beq 
\Delta B_3^{(n)}=\left(\tilde{\alpha} \frac{r}{2
  M}\left(\kappa^{(n)}\right)^3-\tilde{\beta} \frac{H_1}{M^2} 
\left(\kappa^{(n)}\right)^4\right)~.  
\eeq 
If we have renormalized to the experimental binding energy of the
$m$th bound state at leading order we want to preserve that
experimental value at NLO, and so we demand
$\Delta B_3^{(m)}=0$. This fixes the strength of $H_1$ to be
\beq
H_1=\frac{\tilde{\alpha} r M}{2\tilde{\beta}}\frac{1}{\kappa^{(m)}}~, \eeq and the
shift in the binding energy of the $n$th state therefore becomes 
\beq
\frac{\Delta  B_3^{(n)}}{B_3^{(n)}}=\tilde{\alpha} \frac{\kappa^{(n)} r}{2}
\left(1-\frac{\kappa^{(n)}}{\kappa^{(m)}}\right)~.
\label{eq:NLOshift}
\eeq

Parametrically, the fractional change in $B_3^{(n)}$ is of order
$\kappa^{(n)} r$, as expected. However, the number in brackets in
Eq.~(\ref{eq:NLOshift}) is not necessarily of order 1, since the ratios
of binding momenta are large. Because of the discrete scaling symmetry
of the three-nucleon spectrum which is realized in our LO \eftnopi
calculation we have
\begin{equation}
\frac{\kappa^{(n)}}{\kappa^{(m)}}=(22.7)^{m-n}.
\end{equation}
If the state which has been used as a renormalization point
($m$) is more weakly bound than the states for which we are making a
prediction ($n$) we have $m > n$, and so the NLO shift will
be amplified by a factor of $(22.7)^{m-n}$. This to some 
extent explains the rather large shift in $B_3^{(0)}$ when we renormalize to
$B_3^{(1)}$.
Conversely, if the state used for renormalization is more strongly bound
then the state under consideration, and so $m < n$, the NLO shift will be
\begin{equation}
\frac{\Delta B_3^{(n)}}{B_3^{(n)}} \approx \tilde{\alpha} \frac{\kappa^{(n)} r}{2},
\end{equation}
in accordance with naive dimensional analysis.

Equation (\ref{eq:NLOshift}) predicts that \eftnopi should converge
more smoothly for states which are less bound than the state which has
been used for renormalization.  So we now turn our attention to the
convergence of predictions for binding energies of excited states
of the triton, where we expect \eftnopi to work much better than it
does for the ground state. Since we used $B_3^{(1)}$ as input, the
simplest observable we can test this hypothesis on is the ratio
$B_3^{(1)}/B_3^{(2)}$.  (Later we will examine the doublet $nd$
scattering length, $a_{nd}^{1/2}$, but there we do not have $\chi$EFT
data with which to compare.)  Using Eq.~(\ref{eq:NLOshift}), and
remembering that we are effectively choosing $H_1$ such that $\Delta
B_3^{(1)}=0$ , we expect \bea
\label{eq:scaling_error}
\frac{B_3^{(1)}}{B_3^{(2)}}\approx 515\times\left(1 -
\tilde{\alpha} \frac{\kappa^{(2)} r}{2}+\ldots\right)~, 
\eea 
with $\kappa^{(2)}$
the binding momentum of the second excited state.  Since $\kappa^{(2)}
r/2 < 0.1$\%, this suggests that $B_3^{(1)}/B_3^{(2)}$ will change by
less than 0.1\% from LO to N$^2$LO.

\renewcommand{\arraystretch}{1.2}
\begin{table}[t]
\begin{center}
\begin{tabular}{|c|c|c|c|c|}
\hline\hline
        &$B_3^{(0)}$ [MeV]&$B_3^{(0)}/B_3^{(1)}$ & $B_3^{(2)}$ [MeV]&  $B_3^{(1)}/B_3^{(2)}$\\
\hline
 LO     & 3.579 & 515    &$1.349 \cdot10^{-5}$ & 515.0 \\
 NLO    & 3.369 & 485    &$1.350 \cdot10^{-5}$ & 514.8 \\
 N$^2$LO& 3.594 & 517    &$1.350 \cdot10^{-5}$ & 514.8   \\
\hline\hline
$\chi$EFT& 3.774 & 543 &$1.329 \cdot10^{-5}$& 523.1\\  
\hline\hline
\end{tabular}
\vspace{0.3cm}
\caption{\label{tab1} Binding energies $B_3^{(0)}$ and $B_3^{(2)}$ and
  ratios
  $B_3^{(0)}/B_3^{(1)}$ and $B_3^{(1)}/B_3^{(2)}$ at the critical pion
  mass.  The three-body input parameter has been adjusted such that
  the $\chi$EFT result for the first excited triton state
  $B_3^{(1)}=0.00695$~MeV (at $m_\pi=m_\pi^{crit}$) is reproduced.  The
  first three rows show the results in the pionless EFT at LO, NLO,
  and N$^2$LO, while the last row gives the $\chi$EFT result from
  Ref.~\cite{Epelbaum:2006jc}.}
\end{center}
\end{table}

In Table~\ref{tab1} we show the ratios of the first three Efimov
states extracted from the $\chi$EFT and pionless EFT calculations. 
In Ref.~\cite{Epelbaum:2006jc} values of 542.9 and 523.1 were
given for $B_3^{(0)}/B_3^{(1)}$ and $B_3^{(1)}/B_3^{(2)}$
respectively.  Given Eq.~(\ref{eq:NLOshift}) we expect that \eftnopi
converges slowly for binding momenta on the order of $\kappa^{(0)}$,
so we do not anticipate reproduction of $B_3^{(0)}/B_3^{(1)}$ with accuracy
better than a few per cent at N$^2$LO. This expectation is borne out 
by the results of Table~\ref{tab1}.

In contrast, range corrections are miniscule in the case of 
$B_3^{(1)}/B_3^{(2)}$. We have calculated NLO and N$^2$LO $\kappa r$
corrections to this ratio at $m_\pi=m_\pi^{crit}$. We find a shift at
NLO in the ratio that is numerically significant, and on the order of
the expected 0.1\%. No N$^2$LO shift in $B_3^{(1)}/B_3^{(2)}$ is 
seen within our numerical accuracy.  We summarize this result as:
\begin{equation}
\frac{B_3^{(1)}}{B_3^{(2)}}=514.8 \pm 0.02 \pm 0.0004,
\label{eq:EFTpislashratio}
\end{equation}
where the first error is from the numerical precision of our
calculation, and the second is from higher-order terms in the \eftnopi
expansion. 
We note that the $\chi$EFT
calculation of this quantity is numerically quite delicate, since it
requires a momentum mesh that bridges scales ranging from
$\Lambda=540$ MeV (the cutoff in the chiral potential) to the binding
momentum $\kappa^{(2)} \approx 130$ keV---more than three
orders-of-magnitude difference. Indeed, reexamination 
of the calculation of the second excited-state
energy in Ref.~\cite{Epelbaum:2006jc} showed that the 
numerical value of 523.1 quoted there is only accurate to two 
significant digits \cite{noggaprivate}.
There is thus no disagreement between our result
(\ref{eq:EFTpislashratio}) and that quoted in
Ref.~\cite{Epelbaum:2006jc}.
This emphasizes the ability of \eftnopi to easily obtain
high-precision results for three-body observables.

Therefore, at the critical pion mass, $B_3^{(1)}/B_3^{(2)}$ differs
from the Efimov prediction (\ref{eq:efimovratio}) by less than
0.1\%. This allows us to predict that the higher excited states will
obey
\begin{equation}
\frac{B_3^{(1)}(\mpic)}{B_3^{(n+1)}(\mpic)}=(515)^n.
\end{equation}
for all $n \geq 1$.  This result, which is a rigorous consequence of
the discrete scaling symmetry of the Efimov spectrum and the scaling of the
corrections to that spectrum, could be a useful constraint on the
infinite tower of excited states if the critical trajectory is ever
realized in a lattice QCD computation.

\subsection{Doublet Scattering Length}
\begin{figure}[tb]
\centerline{\includegraphics*[width=10cm,angle=0,clip=true]{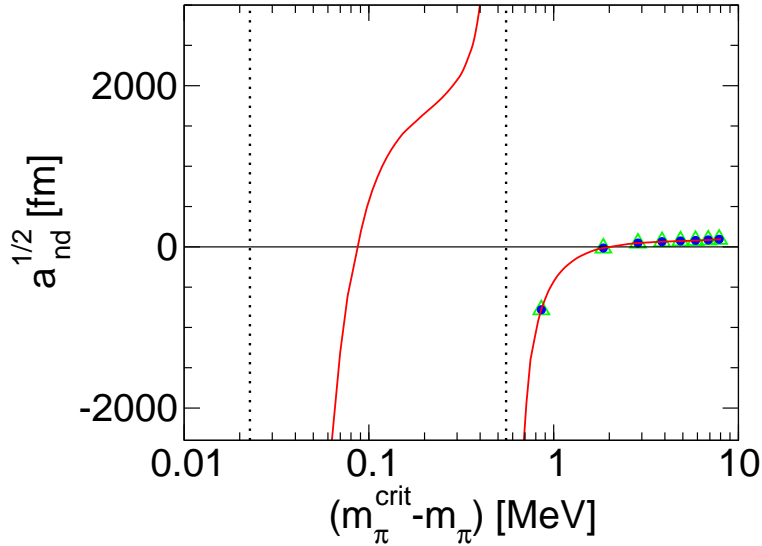}}
\caption{Doublet neutron-deuteron scattering length $a_{nd}^{1/2}$
in the critical region computed in the pionless EFT. The solid line gives 
the LO result, while the triangles and circles show the NLO and 
N$^2$LO results. The dotted lines indicate the 
pion masses at which $a_{nd}^{1/2}$ diverges.
}
\label{fig:aND_crit}
\end{figure}
Scattering observables have not been calculated
in $\chi$EFT in the vicinity of the critical trajectory, but can be
predicted in \eftnopi in a straightforward way. In principle the
subtracted version of Eq.~(\ref{eq:integraleq}) can be solved at any
positive energy to obtain $nd$ phaseshifts. In practice additional
numerical techniques are needed to do this above the three-nucleon
breakup threshold, and at the critical pion mass this coincides with
the $nd$ threshold. However, the technology to deal with this complication
is well known.  Here we examine the $nd$ scattering lengths as a
representative of all $nd$ scattering observables that could be
computed in \eftnopi. The scattering length is measured at zero
momentum. Since we renormalize to a bound state with binding momentum
larger than this we do not anticipate any enhancements of the type
encoded in Eq.~(\ref{eq:NLOshift}).  If one takes into account the
binding momentum of the deuteron, one could argue that $a_{nd}^{1/2}$
involves typical momenta of order $\gamma$, thus the error in the
N$^k$LO \eftnopi prediction for $a_{nd}^{1/2}$ (or for the quartet
scattering length $a_{nd}^{3/2}$) should be, at worst, $(\gamma
r)^{k+1}$.  In Fig.~\ref{fig:aND_crit},
we show the doublet scattering length $a_{nd}^{1/2}$ in the critical
region.  The solid line gives the LO result, while the triangles and
circles show the NLO and N$^2$LO results. The dotted lines indicate
the pion masses at which $a_{nd}^{1/2}$ diverges because the second
and third excited states of the triton appear at the neutron-deuteron
threshold.  These singularities in $a_{nd}^{1/2}(m_\pi)$ are a clear
signature that the limit cycle is approached in the critical
region. Fig.~\ref{fig:aND_crit} shows that, as expected from our
scaling arguments, the higher-order corrections to $a_{nd}^{1/2}$ are
very small and the NLO and N$^2$LO results lie on top of the LO
curve.
Considering the specific case of
the doublet scattering length at
$m_\pi=190$~MeV, we see that
\beq
a_{nd}^{1/2}(m_\pi=190~\text{MeV})=(93.18+0.80+0.14)~\text{fm}~.
\eeq
At this pion mass we are able to match \eftnopi to $B_3^{(1)}$, and we
are also fairly far away from any singular
points in the function $a_{nd}^{1/2}(m_\pi)$. Consequently, the
\eftnopi results follow a natural convergence pattern with the
expansion parameter $\gamma r$, which is $\approx 0.08$ at this $m_\pi$.

\begin{figure}[tb]
\centerline{\includegraphics*[width=10cm,angle=0,clip=true]{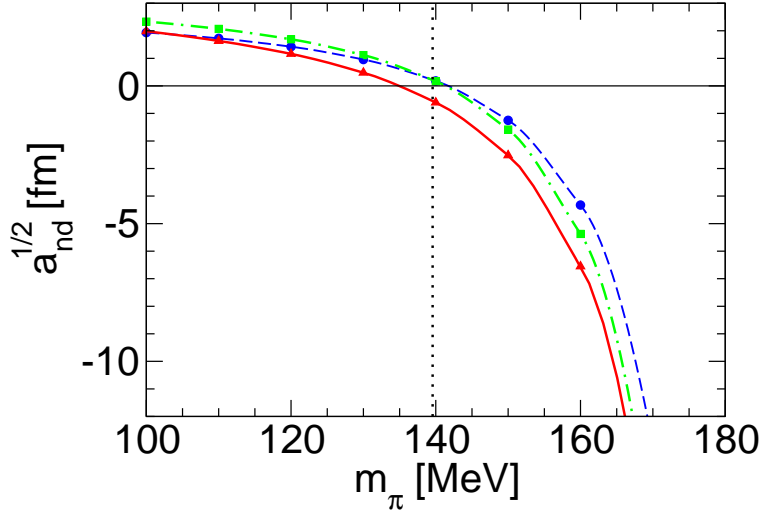}}
\caption{Doublet neutron-deuteron scattering length $a_{nd}^{1/2}$ around
the physical pion mass at LO (dashed), NLO (dash-dotted) and
N$^2$LO (solid). Our results have been connected with splines to
guide the eye. The dotted line gives the physical pion mass.
}
\label{fig:aND_physregion}
\end{figure}

While the Efimov effect leads to dramatic effects around the critical
pion mass, the pionless EFT can also be used for a similar analysis in
the region between $m_\pi^{phys}=139$ MeV and the lower edge of the
critical region.  In this pion-mass domain the triton has no excited states. We
have therefore used the $\chi$EFT results for the pion-mass
dependence of the ground (and only) three-nucleon bound state as the
three-body input to our \eftnopi
computation. We then predict the pion-mass dependence of the doublet
neutron-deuteron scattering length $a_{nd}^{1/2}$.  In
Fig. \ref{fig:aND_physregion} we display \eftnopi results at LO, NLO and
N$^2$LO.
A couple of points in Fig.~\ref{fig:aND_physregion} are particularly
worth noting. First, the convergence pattern of \eftnopi for $m_\pi
\approx 100$ MeV is a little peculiar, with NLO and N$^2$LO
corrections essentially canceling to leave the LO prediction
undisturbed. There is also a cancellation in the NLO correction at
$m_\pi=m_\pi^{phys}$, which leaves the NLO shift in $a_{nd}^{1/2}$
there accidentally close to zero. Presumably this occurs because at
that point the two types of NLO corrections---one proportional to the
binding momentum $\kappa^{(0)}$ and the other proportional to $\gamma
r$, are equal in magnitude but opposite in sign.  This is consistent
with the observed feature that as $m_\pi \rightarrow m_\pi^{crit}$ the
convergence pattern of the EFT becomes more natural, because
$\kappa^{(0)} r$ corrections begin to dominate the $\gamma r$ ones. A
second interesting point is that our results suggest that a small
decrease in the quark masses could lead to a sign change in the $nd$
doublet scattering length.  Perhaps most important, our N$^2$LO
\eftnopi computation allows us to make a firm prediction (within the
scenario under consideration here): as $m_q$ increases above its ``real
world'' value $a_{nd}^{1/2}$ will decrease monotonically until the
first triton excited state appears, something that in the
critical-point realization studied in Ref.~\cite{Epelbaum:2006jc}
happens at $m_\pi \approx 190$ MeV.

\subsection{Quartet Scattering Length}
With results for $a_{nd}^{(1/2)}(m_\pi)$ in hand it is natural to ask
whether \eftnopi can be used to say anything about $nd$
scattering in the other channel, the quartet. In this case the
integral equation describing the scattering is~\cite{STM57,Bedaque:1997qi}:
\begin{equation}
K_{3/2}^{(n)}(q,q';E)=\tilde{\mathcal{Z}}(q,q';E)
+ \mathcal{P}\int_0^\Lambda\hbox{d}q''\,q''^2\tilde{\mathcal{Z}}(q,q'';E)
\tau_t^{(n)}\left(E-\frac{3 q''^2}{4 M}\right)K_{3/2}^{(n)}(q'',q';E)
\label{eq:quartet}
\end{equation}
where 
\begin{equation}
\tilde{\mathcal{Z}}(q,q';E)=\frac{1}{2}\frac{M}{q q'}\log\left(\frac{q^2+q q'
    +q'^2-M E}{q^2-q q'+ q'^2-M E}\right)~.
\label{eq:Zquartet}
\end{equation}
The prefactor in Eq.~(\ref{eq:Zquartet}) is different
from  Eq.~(\ref{eq:Zalphabeta}). 
\begin{figure}[tb]
\centerline{\includegraphics*[width=10cm,angle=0,clip=true]{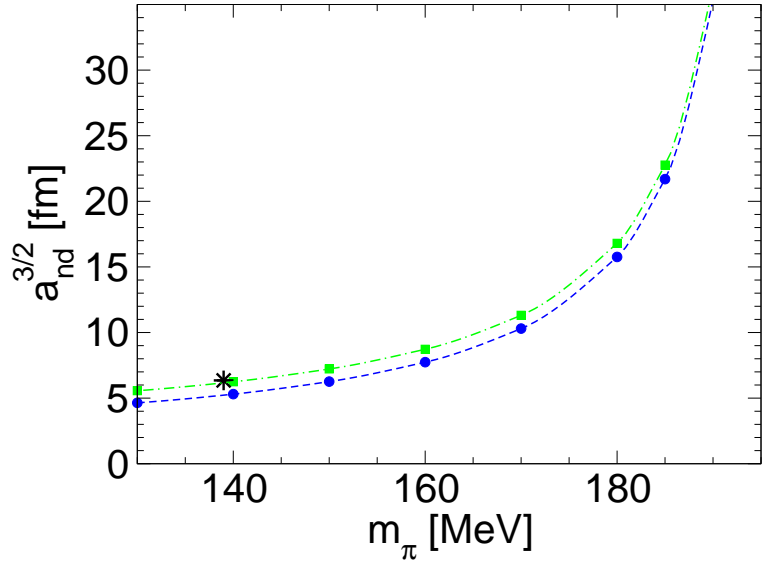}}
\caption{Quartet neutron-deuteron scattering length $a_{nd}^{3/2}$
  around the physical pion mass at LO (dashed) and NLO (dot-dashed). The data
  points have been connected with splines to guide the eye. The dotted
  line gives the physical pion mass. The star indicates the
  experimental value, with the error bar contained within the symbol.}
\label{fig:quarteta}
\end{figure}
This crucial difference
is a consequence of the spin
and isospin coupling in the quartet channel, and
it guarantees that the integral equation (\ref{eq:quartet})
will yield cutoff-independent results without a
subtraction  being necessary. This is a manifestation of the Pauli principle in this
channel, which precludes the appearance of a three-body force without
derivatives. In consequence once $\tau_t^{(n)}$ is defined
Eq.~(\ref{eq:quartet}) is straightforward to solve. Up to 
N$^4$LO the result for $a_{nd}^{3/2}$ is purely a function of
$\gamma_t$ and $r_t$. Performing LO ($n=0$) and NLO ($n=1$)
computations using Eq.~(\ref{eq:quartet}), and then expanding the NLO
result in Taylor series about $r_t=0$, yields the result for this function:
\begin{equation}
a_{nd}^{3/2}=1.1791 \,\gamma_t^{-1} +0.5540(6) \, r_t
+\mathcal{O}\left(\gamma_t r_t^2\right)~.
\label{eq:quarteta}
\end{equation}
(For N$^2$LO computations of this observable at the physical
$\gamma_t$ and $r_t$ see
Refs.~\cite{Bedaque:1997qi,Griesshammer:2004pe}.)  Here the
leading-order coefficient of 1.1791 (accurate to the number of digits
written) was already obtained in Ref.~\cite{STM57}. The computation of
the next-to-leading order coefficient (which has an error of 6 in the
last digit) agrees with results from Efimov~\cite{Efimov91}.  The
result (\ref{eq:quarteta}) translates into the results presented in
Fig.~\ref{fig:quarteta} for $a_{nd}^{3/2}(m_\pi)$.  Numerically we
have, at the physical pion mass,
\begin{equation}
a_{nd}^{3/2}=5.0911 \times (1 + 0.190 \pm 0.139)~{\rm fm}=(6.06 \pm
0.71)~{\rm fm},
\end{equation}
which is consistent with the experimental value of $6.35 \pm 0.02$ fm
for this quantity~\cite{Dilg71}. Indeed, the NLO calculation performed
here agrees with the experimental number at the 5\% level,
which is better than expected from our naive-dimensional-analysis estimate
of the size of the N$^2$LO correction. The accuracy of the \eftnopi
 prediction for $a_{nd}^{3/2}(m_\pi)$ should only improve as
we move from the physical pion mass towards $\mpic$. Note that once
$m_\pi > \mpic$ the quantity $a_{nd}^{3/2}$ cannot be defined, since
the spin-1 $NN$ state becomes unbound, and the spin-0 bound state
obviously cannot be used to produce a total spin of 3/2.

\section{Summary \& Conclusion}
\label{sec-conclusion}
In this paper, we have presented a detailed study of the pion-mass
dependence of three-nucleon observables around the critical pion mass
and around the physical pion mass. We have performed calculations to
N$^2$LO in \eftnopi using the pion-mass dependence of
the two-nucleon effective range parameters and one triton state from
a NLO calculation in $\chi$EFT as input.

In particular, we analyzed the convergence pattern of the
effective-range corrections at NLO and N$^2$LO in the critical region
where $\gamma_s\approx \gamma_t \approx 0$.  We found that the
convergence of the EFT expansion is slow for the triton ground state
but rapid for all remaining states in the three-nucleon Efimov
spectrum. This behavior is expected, since the higher-order
corrections in the critical region scale with powers of $\kappa r$,
where $\kappa$ denotes the typical momentum scale of the observable
under consideration.
Our results demonstrate that the pionless EFT is well
suited to describe low-energy observables in few-body systems with a
large scattering length. For the triton ground state, it appears to be
worthwhile to extend current pionless EFT calculations to N$^3$LO, so
as to
further investigate the convergence pattern.  For the triton excited
states in the critical region, the N$^3$LO corrections will be tiny.
Although an N$^3$LO calculation would require the inclusion of an
additional three-body parameter, the binding energy of the
three-nucleon ground state could be calculated if this parameter was
matched to scattering data or an excited three-body state.

Directly at the critical pion mass, we have performed a detailed
comparison with the triton energies from $\chi$EFT. In the zero-range
limit, the three-body system displays an exact discrete scale
invariance and the ratio of energies of two subsequent Efimov states
is $B_3^{(n)}/B_3^{(n+1)}\equiv\exp(2\pi/s_0)\approx 515.03...$.  For
finite effective range, $r$, there are corrections to this number
proportional to $\kappa^{(n)} r$, where $\kappa^{(n)} \sim
(M B_3^{(n)})^{1/2}$ is the binding energy of the $n$th excited
state. However, the exact ratio $\exp(2\pi/s_0)$ is approached as
$n\to\infty$ since $\kappa^{(n)}\to 0$ in this limit.  For the ratio
of the energies of the first and second excited state, we find
$B_3^{(1)}/B_3^{(2)}=514.8$ when the range corrections are included to
N$^2$LO.  The apparent disagreement of this result with the direct
calculation using the $\chi$EFT potential \cite{Epelbaum:2006jc} was
resolved by taking the reduced accuracy for the excited states in that
calculation into account \cite{noggaprivate}.  In the simpler pionless
EFT, the excited states can be calculated to very high accuracy
without much computational effort and the higher-order corrections are
small. For the more deeply bound states on the other hand, the
convergence of the pionless EFT is slow and high accuracy is possible
in $\chi$EFT.  This calculation provides a prime example of how
$\chi$EFT and \eftnopi complement each other very well.

We have also studied the pion-mass dependence of three-nucleon
scattering observables. 
Directly at the physical pion mass, the $nd$ doublet scattering length
$a_{nd}^{1/2}$ 
is unnaturally small. As a consequence, small variations in the
pion mass lead to significant changes in $a_{nd}^{1/2}$. As the 
pion mass is increased towards the critical value $a_{nd}^{1/2}$ becomes 
very large and negative, jumping to positive infinity when the 
first excited state of the triton appears. This behavior repeats as one 
moves closer to the critical trajectory and more and more excited states 
of the triton appear. It is a signature of the limit cycle 
being  approached. The source of this strong variation
of the doublet scattering length in the critical region 
is therefore understood and well described
by \eftnopi. The pion-mass dependence of the $nd$
quartet scattering length at the physical pion mass is much milder.
As the critical pion mass is approached, it grows and 
eventually becomes infinite
on the critical trajectory. In contrast to the doublet scattering length,
this growth is simply driven by the increase 
in size of the deuteron and has nothing to do with the limit cycle.

Our results will be useful in the context of future Lattice-QCD
calculations of three-nucleon observables. As a first step in this
direction, the $NN$ scattering lengths have recently been extracted
from $NN$ correlators computed at quark masses corresponding to pion
masses between 350 and 590 MeV~\cite{Beane:2006mx}.
Thanks to continuing advances in computer power much progress in this
direction can be expected over the next couple of years. Lattice
computations of $NNN$ correlators are a high priority since they will
provide access to three-nucleon forces directly from QCD, but
algorithmic and theoretical advances will be required in order for
such calculations to become a reality.  One important ingredient in
such calculations will be extrapolations of three-nucleon observables
as a function of pion mass. In particular, a precise understanding of
the behavior of three-nucleon observables in the critical region will
be indispensable in making contact with the physical pion mass.  In this
paper we have investigated the physics involved in such
extrapolations and demonstrated the complementarity of $\chi$EFT
and \eftnopi calculations in their development.

Finally, the possible limit cycle in QCD is also very interesting in its
own right. First signatures of such a limit cycle have recently been
seen in an experiment with cold atoms \cite{Grimm06}, and it would be 
interesting to observe a
limit cycle in lattice simulations of the three-nucleon system 
with pion masses around 200 MeV \cite{Wilson:2004de}. 
The pionless effective theory can be understood as an expansion around
this limit cycle and will be instrumental in interpreting the lattice
results. In this paper, we have demonstrated that the required higher-order
calculations are feasible and converge well.

\begin{acknowledgments}
We thank Evgeny Epelbaum for providing us with his results for the pion-mass
dependence of the effective ranges and U.-G.~Mei\ss ner for a careful
reading of the manuscript. This work was supported by the
Department of Energy under grant DE-FG02-93ER40756, by an Ohio
University postdoctoral fellowship, by the EU I3HP
\lq\lq Study of Strongly Interacting Matter'' under contract number
RII3-CT-2004-506078, by the DFG through funds provided
to the SFB/\-TR 16 \lq\lq Subnuclear structure of matter'', and by the
BMBF under contract number 06BN411.
\end{acknowledgments}

 
\end{document}